\documentclass{ifacconf}

\usepackage{graphicx}      
\usepackage{natbib}        
\usepackage{amsmath} 
\usepackage{amssymb}  
\usepackage{listings}
\usepackage{epstopdf}
\usepackage{algorithmic,algorithm}
\usepackage{color}

\newcommand{\mR}{{\mathbb R}}

\newcommand{\mZ}{\mathbb Z}

\newcommand{\Hb}{\mathbf{H}}

\newcommand{\E}{{\mathbb E}}

\newcommand{\rank}{{\rm rank}}

\newcommand{\bmat}{\begin{bmatrix}}
\newcommand{\emat}{\end{bmatrix}}

\begin{document}
\begin{frontmatter}

\title{Modeling and Identification of Low Rank Vector Processes}


\author[First]{Giorgio Picci}
\author[Second]{Wenqi Cao}
\author[Third]{Anders Lindquist}

\address[First]{Department of Information Engineering, University of Padova, Italy. (e-mail: picci@dei.unipd.it)}
\address[Second]{Department of Automation, Shanghai Jiao Tong University, Shanghai,
China. (e-mail: wenqicao@sjtu.edu.cn).}
\address[Third]{Department of Automation and School of Mathematical Sciences, Shanghai Jiao Tong University, Shanghai, China. (e-mail: alq@math.kth.se)}

\begin{abstract}                
We study modeling and identification of processes with a spectral density matrix of  low rank. Equivalently, we consider processes having an innovation of reduced dimension for which Prediction Error Methods (PEM) algorithms are not directly applicable. We show that these processes admit a special feedback structure with a deterministic feedback channel which  can be used to split the identification in two steps, one of which can be  based on standard algorithms while the other is based on a deterministic least squares fit.
\end{abstract}

\begin{keyword}
Multivariable system identification, low-rank process identification, feedback representation, rank-reduced output noise.
\end{keyword}

\end{frontmatter}
\section{Introduction}
Quite often in the  identification   of large-scale time series one has to deal with  {\em low rank} signals which in general,   have a rank deficient spectral density. These may arise in diverse   areas such as economics, networked systems, neuroscience and so on. \\
Suppose we want to identify  an $(m+p)$-dimensional vector time series $y$ which is weakly stationary, p.n.d. with zero mean and a rational  spectral density $\Phi$ of rank $m$.  This  spectral density  can always be written in factorized form
\begin{equation}\label{sfd}
  \Phi(e^{i\theta}) = W(e^{i\theta})W(e^{-i\theta})^{\top},
\end{equation}
with $W$  an  $(m+p)\times m$ full rank causal rational spectral factor. This spectral rank deficiency case is called \emph{reduced-rank spectra} and  $y$ a \emph{sparse (or singular) signal} in some literature. Researchers discuss singular time series from different points of view. Singular   autoregressive moving average (ARMA) models are discussed in  \cite{Deistler19} or for  factor models see   \cite{DeistlerEJC}; for state space model see  \cite{CLPcdc20}. The identification  of singular models is in particular addressed in \cite{VanDenHof17}, \cite{BLM19}. \cite{VanDenHof17}  proposes  a Prediction Error Method (PEM) identification
of singular time series with reduced-rank output noise. \cite{BLM19} studies the identification of singular vector autoregressive (VAR) models with singular square transfer matrices. In \cite{GLsampling} it was  shown   that there are deterministic relations between the entries of a singular process $y(t)$ while \cite{CLPcdc20} made these deterministic relations specified in a feedback model.

 Let
\begin{equation}\label{ysplit}
    y(t):=\bmat y_1(t) \\ y_2(t) \emat,
\end{equation}
where $y_1(t)$, $y_2(t)$ are jointly stationary of dimension $m$ and $p$. By properly rearranging the  components of $y$, we may assume that $y_1(t)$ is a process of full rank $m$.  Then
\begin{equation}
\label{Phidecomp}
\Phi(z)=\begin{bmatrix}
\Phi_{11}(z)&\Phi_{12}(z)\\ \Phi_{21}(z)&\Phi_{22}(z)
\end{bmatrix}.
\end{equation}
where $\Phi_{11}(z)$ is full rank.
In this paper, we shall show that the  low rank structure implies a deterministic relation between the variables  $y_1(t)$ and $y_2(t)$ which is slightly different from that in \cite{CLPcdc20}. We show that this structure is natural and  helps in the identification of low rank vector processes.

The structure of this paper is as  follows. In Section~\ref{secFB} we introduce feedback models for low-rank processes,  and prove the  existence of a deterministic dynamical relation which reveals the special structure of these processes. In Section~\ref{secIden} we exploit  the special feedback structure for  identification of  the transfer functions of the white noise representation models.  The identification of processes with an external measurable input   is considered in Section~\ref{secIdenExInput}. Several simulation examples are reported in Section~\ref{secExamples}. Finally, we give some conclusions in Section~\ref{secCon}.

\section{Feedback models of stationary processes }\label{secFB}
\begin{figure}[h]
      \centering
      \includegraphics[scale=0.45]{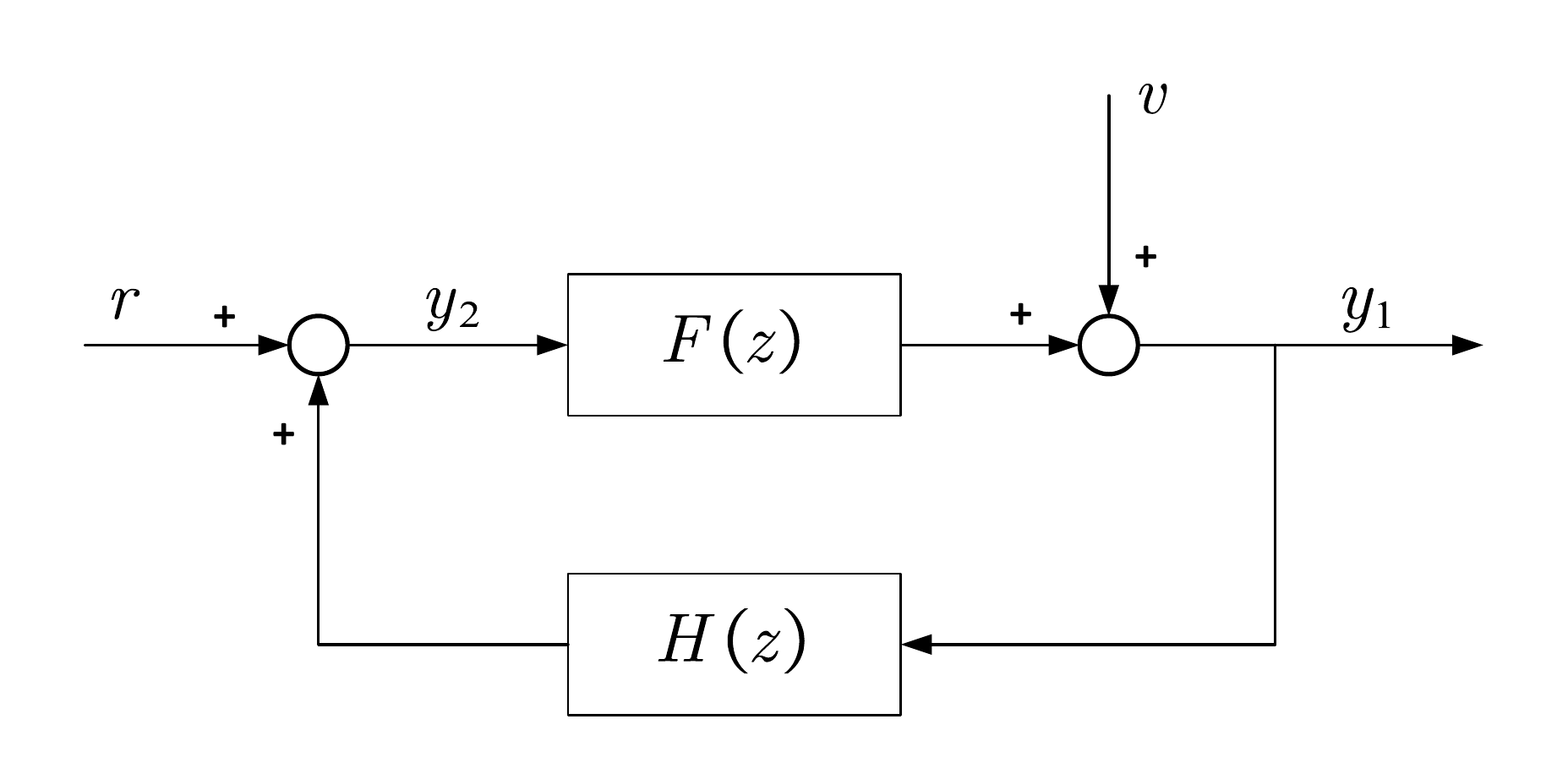}
      \caption{Block diagram illustrating a feedback model}
      \label{FigFH}
\end{figure}

In this section,  we shall first  review  the definition and some properties of general feedback models. Then we will derive a special feedback model for low-rank processes and prove the existence of a deterministic relation between $y_1(t)$ and $y_2(t)$.
\begin{defn}[Feedback Model]
A Feedback model of the joint process $y(t):= \bmat y_1(t) & y_2(t)\emat^{\top}$ of dimension $m+p$ is a pair of equations
    \begin{subequations}\label{fbmodel}
    \begin{eqnarray}
        \label{fby}
        &&y_1(t) = F(z)y_2(t) + v(t),\\
        \label{fbu}
         &&y_2(t) = H(z)y_1(t) + r(t), \quad t\in \mZ
    \end{eqnarray}
    \end{subequations}
satisfying the following conditions:
  \begin{itemize}
   \item $v$ and $r$ are jointly stationary uncorrelated processes called the {\em modeling error} and the {\em input noise};
     \item $F(z)$ and $H(z)$ are $m\times p$, $p\times m$ causal transfer function matrices;
      \item the closed loop system mapping   $[v, r]^{\top}$ to $[y_1, y_2]^{\top}$ is well-posed and {\em internally stable };
        \end{itemize}
\end{defn}
In \eqref{fbmodel} $z$ is the one step ahead shift operator acting as: $z y(t) = y(t+1)$. The block diagram illustrating a feedback representation is shown in Fig.~\ref{FigFH}. Note that the transfer functions  $F(z)$ and $H(z)$ are in general not stable, but the overall feedback configuration needs to be internally stable. In the sequel, we shall often suppress the argument $z$ whenever there is no risk of misunderstanding.
It can be shown that feedback representations of p.n.d. jointly stationary processes always exist. Let $\Hb_t^-(y_1)$ be the closed span of the past components $\{ y_{11}(\tau),  \dots, y_{1m}(\tau)\}\mid \tau < t\}$ of the vector process $y_1$ in the Hilbert space of random variables, and let $\Hb_t^-(y_2)$ be defined likewise in terms of $\{ y_{21}(\tau),  y_{22}(\tau), \dots,  y_{2p}(\tau)\mid \tau < t\}$.
A representation similar to  \eqref{fbmodel} may be gotten from the formulas for causal Wiener filters    expressing both $y_1(t)$ and $y_2(t)$ as a sum of the best linear estimate based on the past of the other process plus an error term
 \begin{subequations}\label{Wiener}
        \begin{align}
    y_1(t) &= \mathbb{E}\{y_1(t)\mid \Hb_t^-(y_2)\} + v(t), \\
    y_2(t) &= \mathbb{E}\{y_2(t)\mid \Hb^-_{t}(y_1)\} + r(t).
\end{align}
\end{subequations}
For a processes  with a rational spectral density the Wiener predictors can be expressed in terms of causal rational transfer functions $F(z)$ and $H(z)$ as in Fig \ref{FigFH}.
Although the errors $v$ and $r$ obtained by the procedure \eqref{Wiener} may be correlated, one can show that there exist feedback model representations where they are uncorrelated.

\begin{thm}
    \label{lem1}
        The transfer function matrix $T(z)$ from $\begin{bmatrix}v\\ r\end{bmatrix}$ to $\begin{bmatrix} y_1\\ y_2\end{bmatrix}$ of the feedback model is given by
\begin{subequations}\label{TPQ}
\begin{equation}  \label{T}
T(z)=\begin{bmatrix}P(z)&P(z)F(z)\\Q(z)H(z)&Q(z)\end{bmatrix},
\end{equation}
with
\begin{equation}  \label{PQ}
\begin{split}
P(z)&=(I - F(z)H(z))^{-1},    \\
Q(z)&= (I - H(z)F(z))^{-1}
\end{split}
\end{equation}
\end{subequations}
where the inverses exist. Moreover, $T(z)$  is a full rank (invertible a.e.) and (strictly) stable function which yields
\begin{equation} \label{Phi}
\Phi(z)=T(z)\begin{bmatrix}\Phi_v(z)&0\\0&\Phi_r(z)\end{bmatrix}T(z)^*,
\end{equation}
where $\Phi_v(z)$ and $\Phi_r(z)$ are the spectral densities of $v$ and $r$, respectively, and $^*$ denotes transpose conjugate.
\end{thm}

\begin{pf}
The  feedback system in Fig.~\ref{FigFH} must be internally stable since the stationary processes $v$ and $r$ produce stationary processes $y$ and $u$ of finite variance. Hence $T(z)$ is (strictly) stable.
From (4) we have
\begin{displaymath}
 \begin{bmatrix}y\\u\end{bmatrix}=\begin{bmatrix}0 & F(z)\\H(z) & 0\end{bmatrix}\begin{bmatrix}y\\u\end{bmatrix}+ \begin{bmatrix}v\\ r\end{bmatrix}
\end{displaymath}
         and therefore
\begin{displaymath}
N(z)\begin{bmatrix}y\\u\end{bmatrix}=\begin{bmatrix}v\\ r\end{bmatrix} ,
\end{displaymath}
               where
 \begin{displaymath}
 N(z) :=\begin{bmatrix}I & -F(z)\\-H(z) & I\end{bmatrix}.
\end{displaymath}
Now the transfer function  $ I - H(z)F(z) $ must be invertible by well-posedness of the feedback system and consequently, $N(z)$ is invertible, while  a straightforward calculation shows that $T(z)N(z)=I$ and hence
$T(z) = N(z)^{-1}$, as claimed. Then \eqref{Phi} is immediate. \hfill$\Box$
\end{pf}

Since $T(e^{i\theta})$ has full rank a.e., $\Phi$ is rank deficient if and only if at least one of $\Phi_v$ or $\Phi_r$ is. 

\begin{lem}\label{lem2}
Suppose $(F\Phi_rF^*+\Phi_v)$ is positive definite a.e. on the imaginary axis. Then
\begin{equation} \label{H}
H=\Phi_{21}\Phi_{11}^{-1} - \Phi_rF^*(\Phi_v+F\Phi_rF^*)^{-1}(I-FH),
\end{equation}
that is
\begin{equation}\label{DetH}
H=\Phi_{21}\Phi_{11}^{-1}
\end{equation}
if and only if $\Phi_r \equiv 0$.
\end{lem}
\begin{pf}
From \eqref{TPQ} and \eqref{Phi}, we have
\begin{align*}
   \Phi_{21}&= Q(H\Phi_v +\Phi_r F^*)P^*=QH\Phi_vP^* + Q\Phi_rF^*P^*,  \\
   \Phi_{11}  &=  P(\Phi_v+F\Phi_rF^*)P^* ,
\end{align*}
and using the easily verified relations
\begin{displaymath}
PF=FQ, \quad HP=QH.
\end{displaymath}
we get
$$
 \Phi_{21}=HP\Phi_vP^* +Q\Phi_rF^*P^*.
 $$
 Adding and subtracting the term $HP F\Phi_rF^*P^*$ we end up with
\begin{align*}
 \Phi_{21}  &=  H \Phi_{11}+ (Q-QHF)\Phi_rF^*P^*\\
 & =H \Phi_{11} + \Phi_rF^*P^*
 \end{align*}
 since $Q-QHF=I$. Then  \eqref{DetH} follows if and only if  $\Phi_r=0$ since $P$ is invertible  and $F$ times a spectral density can be identically zero only if the spectral density is zero (as otherwise  this would imply that the output process of a filter with stochastic input  would have to be  orthogonal to the input).
 \hfill$\Box$\end{pf}

In the following we specialize to feedback models of rank deficient processes. We shall show that there are  feedback model representations  where the feedback channel is described by  a {\em deterministic relation} between $y_1$ and $y_2$.

\begin{thm}\label{thmspecialfb}
Let $y$ be an $(m+p)$-dimensional process  of rank $m$. Any full rank $m$-dimensional subvector process $y_1$ of $y$ can be represented by a feedback scheme of the form
\begin{subequations}\label{fbr0}
 \begin{eqnarray}
    y_1 &=& F(z)y_2+ v, \\
    y_2 &=& H(z)y_1.
 \end{eqnarray}
  \end{subequations}
   where the input noise $v$ is of full rank $m$.
\end{thm}

\begin{pf}
Recall that $n$-tuples of real rational functions form a vector space $\mR^n(z)$ where the rank of a rational matrix is the rank almost everywhere.

The claim is equivalent to the two statements\\
1. If we have the structure \eqref{fbr0}, i.e. $\Phi_r \equiv 0$;  then $y_1$ is   of full rank $m= \rank (\Phi)$.\\
2. Conversely if $y_1$ is   of full rank $m= \rank (\Phi)$ then $\Phi_r \equiv 0$.

Part 1 follows from Lemma \ref{lem2} since because of \eqref{Phi} then $\Phi_v$ must have rank $m (= \rank (\Phi))$.\\
Part 2 is not so immediate. One way  to show it could be as follows.

Since 	$\Phi(z)$ has rank $m$ a.e. there must be a full rank $p \times (m+p)$ rational matrix which we write in partitioned form, such that
\begin{align*}
[A(z)  B(z)] \Phi(z) &=0\, \Leftrightarrow \, [A(z)  B(z)] \bmat\Phi_{11}(z)\\ \Phi_{21}(z)\emat =0 \\
&\, \Leftrightarrow \, [A(z)  B(z)] \bmat y_1(t) \\ y_2(t)\emat =0
\end{align*}
where the last formula has the usual interpretation.

We claim that $B(z)$ must be of full rank $p$. One can prove this using the invertibility of $\Phi_{11}(z)$. Just  multiply from the left the second relation by any  $p$-dimensional row vector $a(z)$ such that $a(z)B(z)=0$.  This would imply that also  $a(z)A(z)\Phi_{11}(z)=0$ which is impossible since $\Phi_{11}(z)$   is full rank and   $a(z)B(z)$ cannot be zero as the whole matrix $ [A(z)  B(z)]$ is full rank $p$.  Now  take any nonsingular $p\times p$ rational matrix $M(z)$ and consider instead $M(z)[A(z)  B(z)]$, which  provides    an equivalent relation. By choosing $M(z)=B(z)^{-1}$ we can reduce $B(z)$ to the identity to get
$$
[H(z) \, I\,] \bmat  y_1(t) \\ y_2(t)\emat =0
$$
where  $H(z)$ is a rational matrix function, so that    one gets the deterministic dynamical relation
$$
  y_2(t)= H(z) y_1(t)\,.
$$
Substituting in  the general feedback model one concludes  that $u$ must then be a functional of only the noise $v(t)$ since  $y(t)$ is such. Therefore $r$  is the zero process i.e. $\Phi_r=0$. Hence by Lemma \ref{lem2} we obtain
$H(z)=\Phi_{21}(z) \Phi_{11}(z)^{-1}$.\hfill$\Box$
\end{pf}

\section{Identification of low rank processes}\label{secIden}
Suppose we want to identify, say by a PEM method,  a  low rank model of an $(m+p)$-dimensional time series,
\begin{equation}
 y(t) = W(z)e(t),
\end{equation}
with $e(t)$ an $m$-dimensional white noise of full rank. Assume  $y_1$ and $y_2$ are described by the special feedback model \eqref{fbr0} and introduce the transfer functions
\begin{equation}
   y(t)= \bmat y_1(t) \\ y_2(t) \emat := \bmat W_1(z)\\ W_2(z) \emat e(t),
 \end{equation}
so that  $W_2(z)=H(z)W_1(z)$. Since $y_1$ (and $W_1$) is full rank, we can identify an ARMA innovation model  for $y_1$ based only on observations of $y_1(t)$ on some large enough interval. Next, since the relation between $y_2$ and $y_1$ is completely deterministic (see \eqref{fbr0}) we can  identify  $H(z)$ by imposing  a deterministic transfer function model   to the observed data, written  $ A(z^{-1}) y_2(t) - B(z^{-1})y_1(t)=0,\, t=1,\ldots,N$ (the minus sign is for convenience) where  $A(z^{-1})$ and $B(z^{-1})$ are matrix polynomials in the delay variable $z^{-1}$ of dimension $p\times p$ and $p\times m$  such that
  $$
  H(z)= A(z^{-1})^{-1}B(z^{-1})
  $$
is causal. One can always choose $A(z^{-1})$ monic and $B(z^{-1})$ (possibly with the zero degree coefficient $B_0=0$) so  that the transfer function corresponds to the model
\begin{equation} \label{procedure1}
y_2(t)= \sum_{k=1}^q A_k y_2(t-k) + \sum_{k=0}^r B_ky_1(t-k), \qquad t=1,\ldots,N,
\end{equation}
 where we have been writing  $A(z^{-1}) = I -\sum_{k=1}^{q}A_k z^{-k}$ and   $B(z^{-1})= \sum_{k=0}^{r}B_k z^{-k}$. The above equation  involves  delayed components of the  observed trajectory data of $y$. The  coefficients can then be estimated  by  solving a deterministic overdetermined  linear system by least squares.

Since the procedure above ignores the structure of the first equation in model \eqref{fbr0}, we need to work with a model involving both transfer functions $F$ and $H$. The model,   assumed in innovation form (an innovation representation is needed to guarantee  model uniqueness i.e.identifiability),  is
\begin{subequations}
\begin{eqnarray}
   y_1 &=& F(z)y_2+K(z)e,    \label{y1Fy2Ke}         \\  \label{y2Hy1}
   y_2 &=& H(z)y_1.
\end{eqnarray}
\end{subequations}
with $K(z)$  a square  spectral factor representation, i.e. $v(t):=K(z)e(t)$, which we assume   normalized at infinity, i.e. $K(\infty)=I$,   and both  $P(z)K(z)$ and $H(z)P(z)K(z)$   minimum-phase. Note that From \eqref{TPQ} we have
\begin{equation}
\label{ }
\bmat W_1 \\ W_2 \emat =T\bmat K\\ 0 \emat = \bmat PK \\ QHK\emat =\bmat PK \\ HPK\emat\,.
\end{equation}
One may ask how one can   recover the direct transfer function $F(z)$ from the identified $W_1(z)$ and $H(z)$. This would amount to solving for $F$ the relation $W_1=(I-FH)^{-1}K$ which, assuming $H$ is given, contains two unknowns. Hence $F$ is not identifiable by this procedure.

Instead  we can transform  \eqref{y1Fy2Ke} into an ARMAX model by using matrix-fraction descriptions. Although this model has (deterministic) feedback, the Prediction Error method, see    \cite{Ljungbook}, allows  to identify these transfer functions. To avoid bringing in the dynamics of $y_2$, we should impose     $F(z)$ to have at least a unit delay, that is $F(z)=z^{-1}F_1(z)$. Then, in force of  the normalization $K(\infty)=I$, we may  write the transfer function of the one-step predictor (and thus the prediction error) by substituting the one-step delay of the innovation    $e(t)= K(z)^{-1} [y_1(t)-F(z) y_2(t)]$ into
\begin{equation}
\hat y_1(t\mid t-1)= F_1(z) y_2(t-1) + \tilde K(z) e(t-1),
\end{equation}
where $ \tilde K(z):= z(K(z)-I)$. One can do these operations in  terms of matrix fraction descriptions and carry on  the PEM optimization  with respect to  the  coefficients of the matrix polynomials. Note that this procedure works {\em without knowing the dynamics of the "input" $y_2$} (i.e. no need to know $H(z)$). If needed, $H(z)$ can be identified independently as seen in the previous paragraph.

\subsection{Details of the ARMAX  identification }\label{SubARMAX}
To identify $F$ and $K$ we write the equation \eqref{y1Fy2Ke} as an ARMAX model,
\begin{equation} \label{armax}
    A(z^{-1})y_1(t)=B(z^{-1})y_2(t)+C(z^{-1})e(t),
\end{equation}
where $F(z)=A(z^{-1})^{-1}B(z^{-1})$, $K(z)=A(z^{-1})^{-1}C(z^{-1})$ are coprime matrix fraction descriptions with $A$ monic (of course these are  not the same polynomials as in the previous paragraph). Although this model has (deterministic) feedback, the PEM allows us to identify these polynomials (actually to this end we also need some extra information or a suitable procedure to guess the degrees and the structure of the matrix polynomials).
To guarantee  well-posedness of the feedback system either $F(z)$ or $H(z)$ (or both) must have a delay. Assume that  $F(z)$ has at least a unit delay, that is
$$
F(z)=z^{-1}F_1(z)= A(z^{-1})^{-1} [z^{-1}B_{1}(z^{-1})].
$$
 Then, if $C_1(z^{-1})$ is the  remainder after a one-step division of $C$ by $A$, i.e.,
$$
C(z^{-1})= A(z^{-1})+ z^{-1} C_1(z^{-1})\, ,
$$
\eqref{armax} can be written
\begin{equation}
\label{ }
\begin{split}
C(z^{-1})y_1(t)=C_1(z^{-1})y_1(t-1) + B_{1}(z^{-1})y_2(t-1)\\ +C(z^{-1})e(t),
\end{split}
\end{equation}
and consequently
\begin{equation}\label{armax1}
\begin{split}
&C(z^{-1}) \hat y_1(t\mid t-1)\\
&\phantom{xxxx}= C_1(z^{-1})y_1(t-1)  +  B_{1}(z^{-1}) y_2(t-1).
\end{split}
\end{equation}
Then the recursion \eqref{armax1} can be used to compute the prediction error $\varepsilon_1(t\mid t-1)= y_1(t)-\hat{y}_1(t\mid t-1)$. We do not consider  here the difficulties connected to {\em parameter identifiability} of these representations in the vector case, since this is a theme which has been amply discussed in the literature.

\section{Identification of a low rank model with an external input}\label{secIdenExInput}

 Referring to a problem  discussed by \cite{VanDenHof17}, suppose we want to identify a multidimensional system  with an external input $u(t)$, say
\begin{equation}\label{WithInput}
   y= Fu + K e
\end{equation}
where $e$ is a white noise process whose dimension is {\em strictly smaller than the dimension of} $y$ and the input $u$ is completely uncorrelated with $e$. In this case  the  model is called low-rank. \\
When $\dim e =\dim y$ and $K(z)$ is square invertible one could attack the problem by a standard PEM method. The method however runs into difficulties when the noise  is of smaller dimension than $y$ since then the predictor and the prediction  error are not well-defined.

Referring to the general feedback model for the joint process we can always assume $F$ causal and    $K(\infty)$ full rank and  normalized in some way. Consider then the prediction error of $y(t)$ given the past history of $u$. We have
\begin{equation}\label{tildey}
\tilde y(t):= y(t)- \E [y(t) \mid \Hb_t(u) ] = K(z) e(t)
\end{equation}
since by causality of $F(z)$ the  Wiener predictor is exactly $F(z)u(t)$. Hence $\tilde y$ is a low rank time series in the sense described in the previous section (with $W(z)\equiv K(z))$. In principle we could then  use the procedure described above for time series as we could preliminarily estimate $F(z)$  by solving a deterministic regression of $y(t)$ on the past of $u$ and hence get $\tilde y(t)$.

\section{Simulation Examples}\label{secExamples}

 \subsection{Example 1}
As a first example  consider   a two-dimensional process of rank 1 described by
\begin{equation}\label{W1W2}
 y(t)= \bmat W_1(z) \\ W_2(z) \emat e(t)
\end{equation}
 where both $W_1(z)$ and $ W_2(z)$ are causal  and stable rational transfer functions and $e$ is a scalar  white noise of variance $\lambda^2$. By simulation we produce a sample of two-dimensional  data. With these data we shall:
 \begin{itemize}
 \item Identify a model for $y_1$  and compute $H(z)$ according to the first procedure. Compute $W_2$ by using $W_2=HW_1$ and check if it  is identified correctly.
  \item Identify $F$ and $K$ using the ARMAX model with input $y_2$ (second procedure) and do the same for the other component.
   \end{itemize}
   We start by  simulating a two-dimensional process $y(t)$ of rank 1 described by \eqref{W1W2} where  $e$ is a scalar zero mean white noise of variance $\lambda^2= 1$ and choose
\begin{eqnarray}\nonumber
W_1(z) &=& \frac{1}{1-0.2z^{-1}-0.25z^{-2}+0.05z^{-3}}, \\
  \nonumber
  W_2(z) &=& \frac{1}{1-0.6z^{-1}+0.03z^{-2}+0.01z^{-3}},
\end{eqnarray}
which are   causal and stable (in fact minimum phase) rational transfer functions. Note that in this particular example {\em both $y_1$ and  $y_2$ are full rank so that our procedure would work for both.}\\
We generate a two-dimensional  time series of $N=500$ data points $\{\bar y_i(t); t=1,\ldots,N,\, i=1,2 \}$.

\begin{figure}
      \centering
      \includegraphics[scale=0.64]{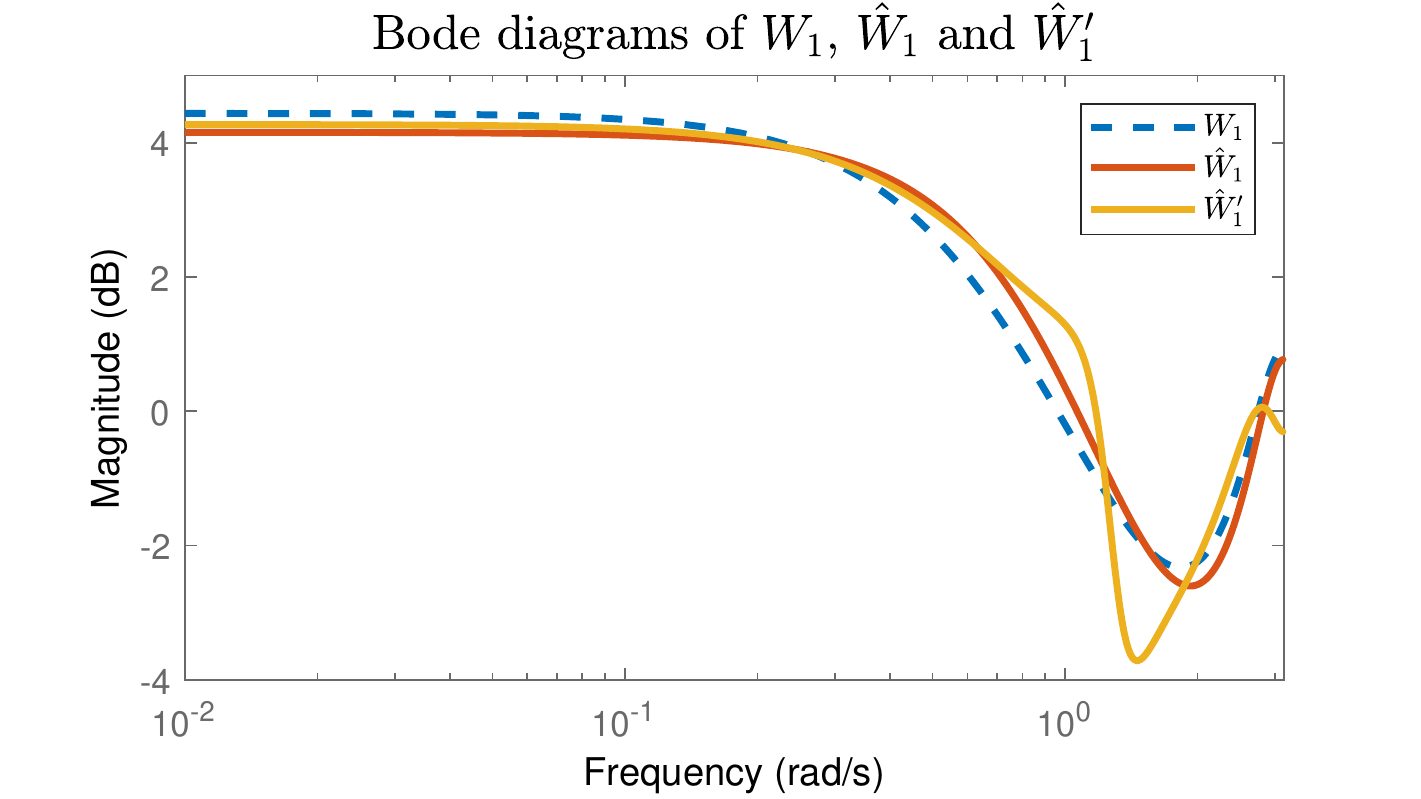}
      \caption{Bode diagrams of $W_1$, $\hat{W}_1$ and $\hat{W}_1'$}
      \label{FigExmp1W1}
\end{figure}

\begin{figure}
      \centering
      \includegraphics[scale=0.64]{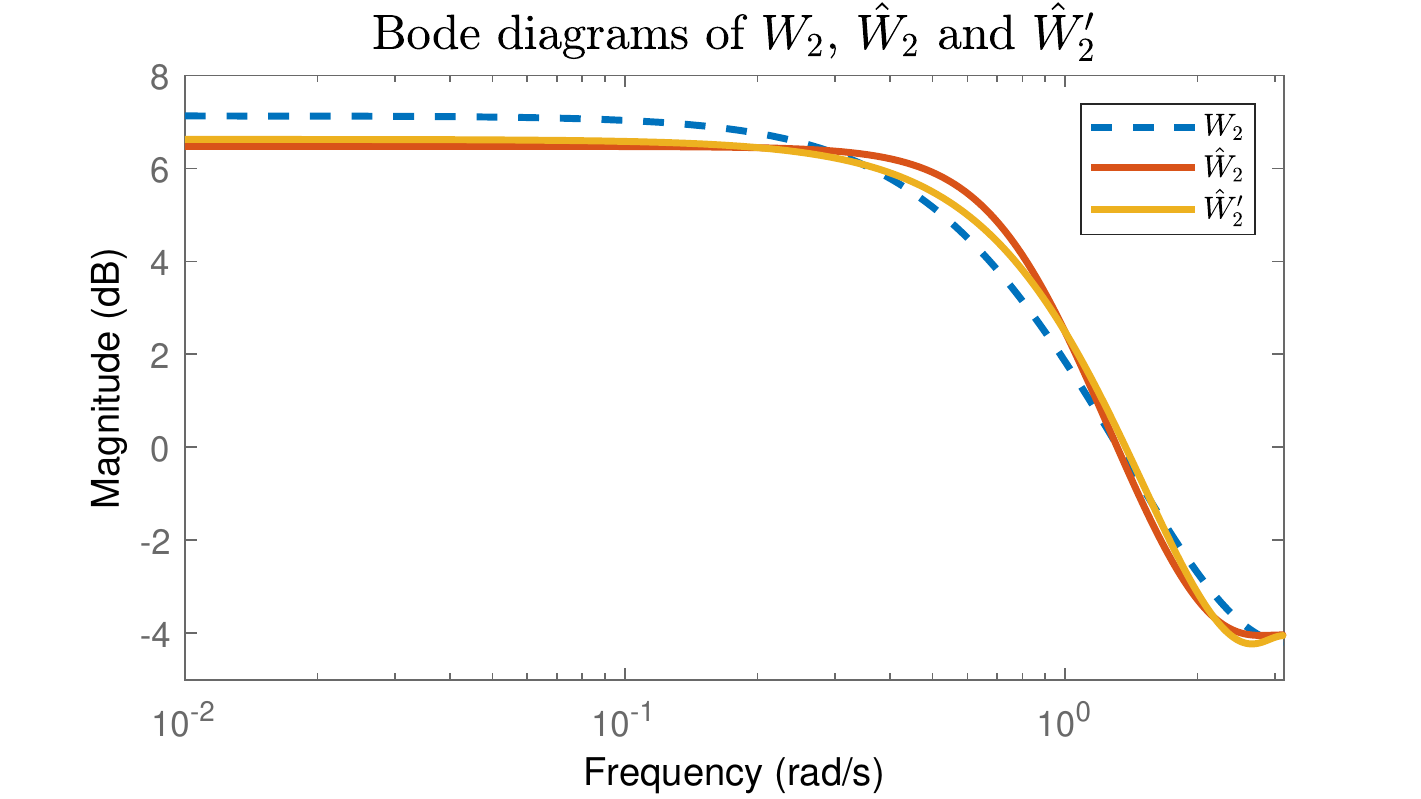}
      \caption{Bode diagrams of $W_2$, $\hat{W}_2$ and $\hat{W}_2'$}
      \label{FigExmp1W2}
\end{figure}

\begin{figure}
      \centering
      \includegraphics[scale=0.64]{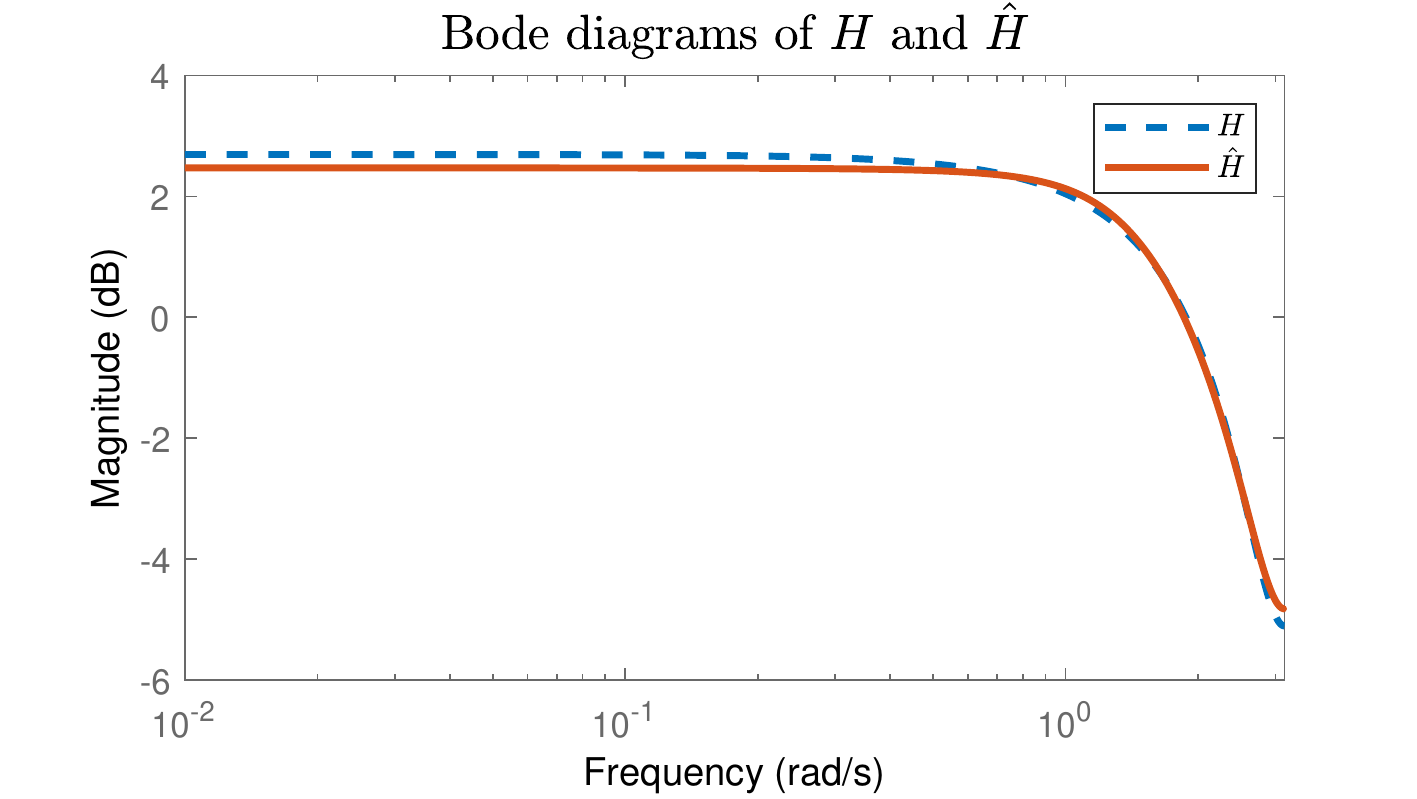}
      \caption{Bode diagrams of $H$, $\hat{H}$}
      \label{FigExmpH}
\end{figure}

\begin{figure}
      \centering
      \includegraphics[scale=0.64]{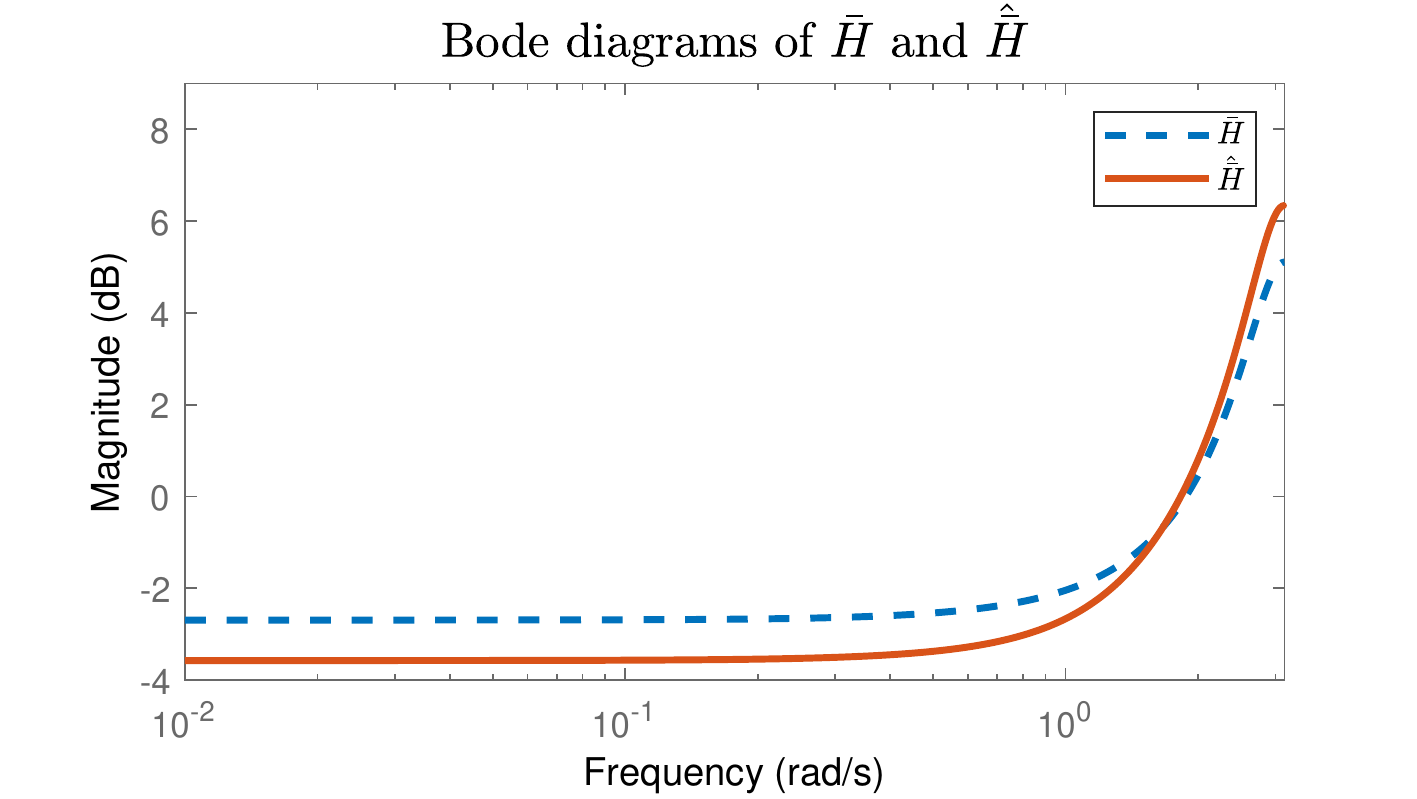}
      \caption{Bode diagrams of ${\bar H}$, $\hat{\bar H}$ }
      \label{FigExmpHbar}
\end{figure}

Since the two AR models of $y_1$ and $y_2$ are  of order 3  (we assume the order  is known) we have to do two AR identification runs in MATLAB  for models of the form
$$
y_i(t) = -\sum_{k=1}^3 a_{i,k} y_i(t-k) + e(t), \qquad t=1, \ldots N,
$$
 for $i=1,2$ to obtain the estimates
$$
 \hat{W_i}=\frac{1}{1+\sum_{k=1}^{3}\hat{a}_{i,k}z^{-k}}.
$$
We get  the following  parameter estimates $\{\hat a_{i,k}\}$ for the two models \begin{equation}\nonumber
\begin{split}
  \hat{a}_{1,1}=-0.2429,~~ \hat{a}_{1,2}=-0.2325,~~ \hat{a}_{1,3}=0.09528;\\
  \hat{a}_{2,1}=-0.6363,~~ \hat{a}_{2,2}=0.03302,~~ \hat{a}_{2,3}=0.07769.
\end{split}
\end{equation}
The Bode graphs of the estimated transfer functions $\hat{W}_{i}$ compared with the true $W_{i}$ are shown in Fig.~\ref{FigExmp1W1} and Fig.~\ref{FigExmp1W2}, where the blue dash lines denote $W_i$, and red line denote $\hat{W}_{i}$.
From the numerical results and graphs we see that the estimated transfer functions are close to the true ones both on parameter values and on the magnitude Bode graphs, which shows that the identification of $W_i$ from AR models works well.\\
Now the theoretical $H(z)$ satisfies the identity
$$
    W_2(z)=H(z)W_1(z),\quad W_1(z)={\bar H}(z)W_2(z),
$$
which implies the theoretical formulas for $H$ and ${\bar H}$:
\begin{equation}\nonumber
 H(z)=\frac{1+0.5z^{-1}}{1+0.1z^{-1}}\quad {\bar H}(z)=\frac{1+0.1z^{-1}}{1+0.5z^{-1}}.
\end{equation}
which are equivalent  to the difference equation
\begin{equation}\nonumber
    (1+0.1z^{-1})y_2(t)- (1+0.5z^{-1})y_1(t)=0,
\end{equation}
that is
$$ y_2(t)=-0.1y_2(t-1)+y_1(t)+0.5y_1(t-1).
$$
 These are just theoretical models which we keep for comparison. Since we don't know the true coefficients  we shall just use the least squares estimates of the second  transfer function  to get
\begin{equation}\nonumber
\begin{split}
   \hat H(z)&=\frac{1+\sum_{k=1}^3 \hat b_{k}z^{-k}}{1+\sum_{k=1}^3 \hat a_{k}z^{-k}}\\
  & = \frac{1+0.2236z^{-1}-0.0124z^{-2}+0.0484z^{-3}}{1-0.1653z^{-1}+0.0973z^{-2}+0.0157z^{-3}}.
\end{split}
\end{equation}
which is a good approximation of the theoretical $H(z)$ as seen in Fig~\ref{FigExmpH}. Using $\hat{H}$ and $\hat{W_1}$, we may calculate an estimate of $W_2$ denoted $\hat{W_2}':=\hat{H}\hat{W_1}$. The Bode graph  of $\hat{W_2}'$ is shown in orange   in Fig.~\ref{FigExmp1W2}.  Results show that, though we don't know the orders of the denominator and numerator of $H$, the Bode graph of $\hat{H}$ fits that of $H$ well. From estimates of $H$ and $W_1$, we may also easily obtain an estimate of $W_2$ which  is as good as the estimate obtained by  by direct identification.\\
By switching the role of the two components  $y_1$ and $y_2$, we may also estimate ${\bar H}(z)$, assumed  of the form
\begin{equation}\nonumber
   {\bar H(z)}=\frac{1+\sum_{k=1}^3 b_{k}z^{-k}}{1+\sum_{k=1}^3 a_{k}z^{-k}},
\end{equation}
and obtain the following estimate,
$$
  \hat{\bar H}(z)=\frac{ 1 - 0.1503 z^{-1} + 0.07048 z^{-2} + 0.005883z^{-3}}{1 + 0.3678 z^{-1} - 0.008278 z^{-2} + 0.03837z^{-3}},
$$
 the compared Bode graphs are shown in Fig.~\ref{FigExmpHbar}.

Next we want to identify $F(z)$ and $K(z)$ in the feedback model. To this purpose we
use the ARMAX identification   algorithm described in subsection \ref{SubARMAX}, referring to a  model \eqref{armax}, with input $y_2$ and output $y_1$.\\
Since we do not know the true orders, we suppose 
\begin{equation}\nonumber
\begin{split}
   A(z^{-1})&= 1+\sum\limits_{k=1}^{3}a_{k}z^{-k}, \\
   B(z^{-1}) &=z^{-1}B_1(z^{-1})=z^{-1}(\sum\limits_{k=0}^{3}b_{k}z^{-k}),\\
   C(z^{-1})&= 1+\sum\limits_{k=1}^{3}c_{k}z^{-k}.
\end{split}
\end{equation}
Note that $A$ should have the same order as $C$, since we have assumed that $K=A^{-1}C$ is normalized at $\infty$. The estimation results are
$$
    \hat{F}=\frac{-0.02217 z^{-1} - 0.02322 z^{-2} - 0.3411 z^{-3} + 0.2154 z^{-4}}{1 + 0.0009619 z^{-1} + 0.04707 z^{-2} + 0.02051 z^{-3} },
$$
$$
    \hat{K}=\frac{1 + 0.2588 z^{-1} + 0.4005 z^{-2} + 0.4596 z^{-3}   }{1 + 0.0009619 z^{-1} + 0.04707 z^{-2} + 0.02051 z^{-3}}.
$$
 With these estimates we  then calculate a corresponding  estimate $\hat{W}'_1$ of $W_1$ by the formula
$$
\hat{W}'_1=(1-\hat{F}\hat{H})^{-1}\hat{K}.
$$
Its Bode graph is the orange line, compared with $W_1$ and $\hat{W}_1$ in Fig.~\ref{FigExmp1W1}.   Since $\hat{W}_1$ has larger orders of both numerator and denominator than those of $W_1$, there is some overfitting and the Bode graph of $\hat{W}_1'$ is not as smooth as those  of $W_1$ and $\hat{W}_1$ in the   high frequency range.

\subsection{Example 2 }
In this subsection and in the next one  we consider    the identification of two-dimensional processes of rank 1 subjected to an external input $u$. We generate  a scalar white noise $u$ independent of $e$ and identify a  2-dimensional process model \eqref{WithInput} as described in the previous section \ref{secIdenExInput}.\\
 In this example  the  true system  is described by
\begin{equation}\label{Ex2Syst}
\begin{split}
   F(z) &=  z^{-1}\bmat 0.3 + 0.7z^{-1}+0.3z^{-2}\\
                 0.15 + 0.9z^{-1}-0.5z^{-2}\emat, \\
   K(z) &= \bmat \frac{1+0.1z^{-1}+0.4z^{-2}}{1+0.3z^{-1}+0.4z^{-2}}\\ \frac{1-0.1z^{-1}+0.4z^{-2} }{1-0.2z^{-1}+0.1z^{-2}} \emat\,.
\end{split}
\end{equation}
We use the same $F$ as in \cite{VanDenHof17} (where it is called $G(q)$). But their $K$ is not normalized, so we use a different one. Both components of our $K(z)$  here are normalized and minimum-phase  so the overall model is an innovation model.

\begin{figure}
      \centering
      \includegraphics[scale=0.6]{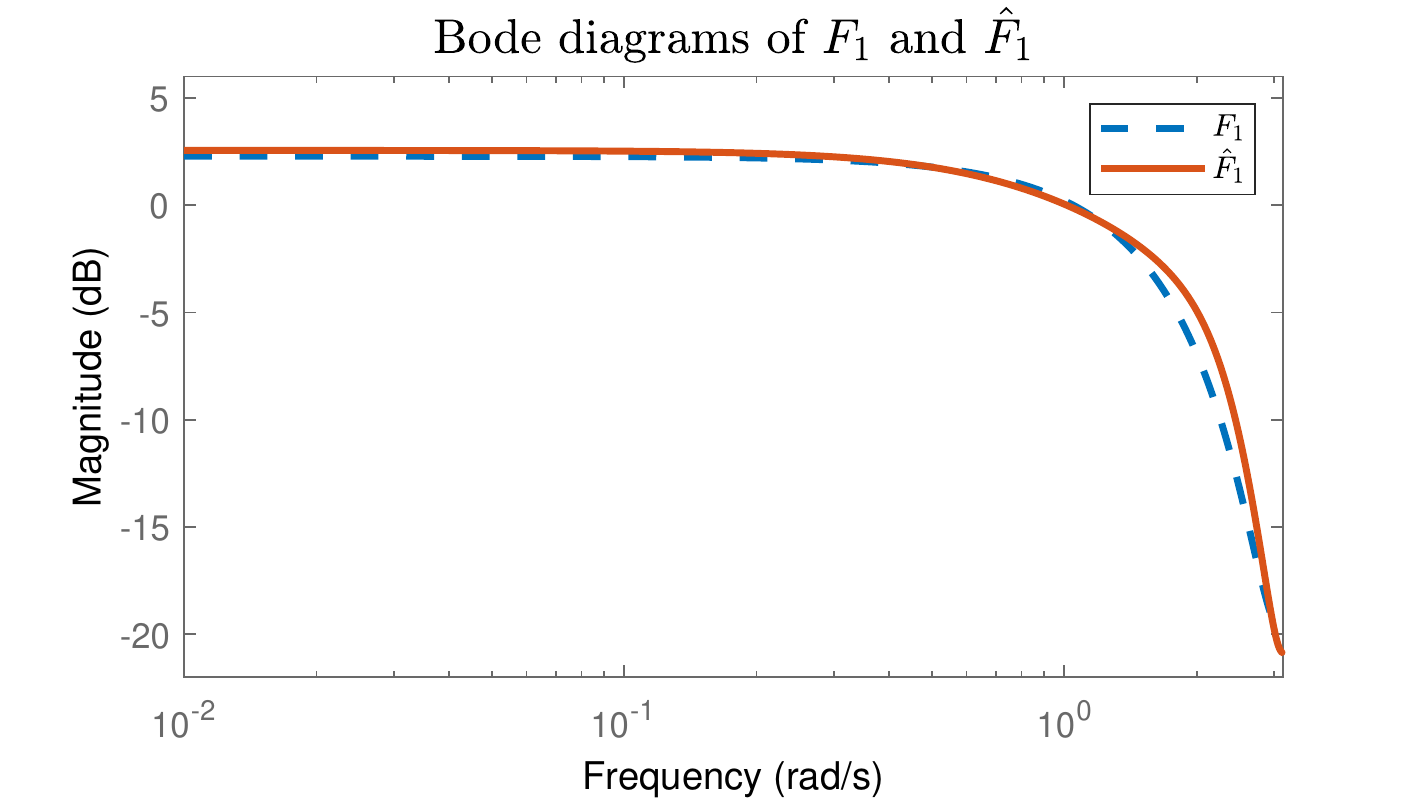}
      \caption{ Bode diagrams of $F_1$ and $\hat{F}_1$ in example 2}
      \label{FigExmp2F1}
\end{figure}

\begin{figure}
      \centering
      \includegraphics[scale=0.6]{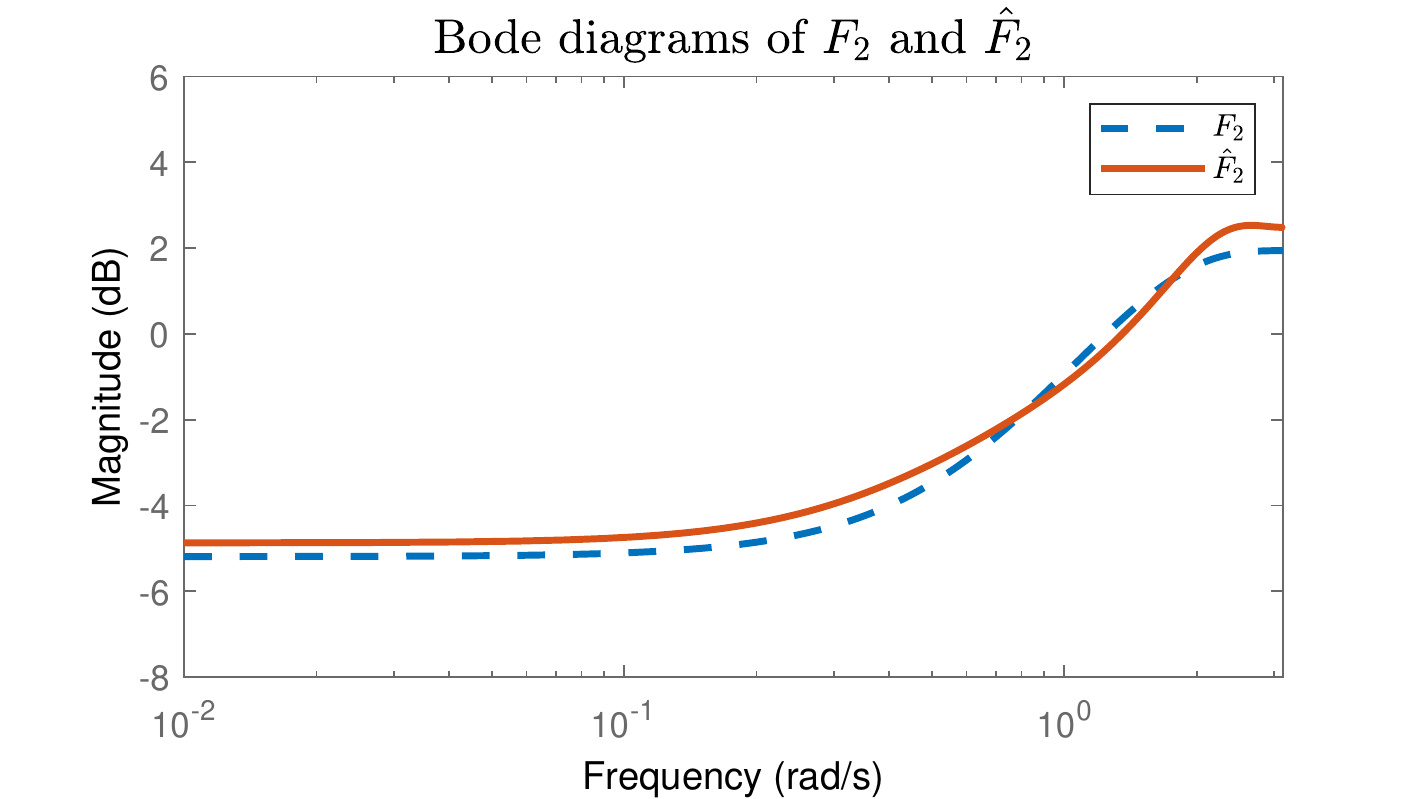}
      \caption{ Bode diagrams of $F_2$ and $\hat{F}_2$ in example 2}
      \label{FigExmp2F2}
\end{figure}

From the model \eqref{Ex2Syst} we generate a two-dimensional  time series of $N=500$ data points $\{\bar y_i(t); t=1,\ldots,N,\, i=1,2 \}$. The simulation  is run with $u$ and $e$ two independent scalar white noises of variances $2$ and $1$.
Of course here  we also measure  the input time series   $u$. First, we estimate $F(z)$ by fitting the deterministic relations
\begin{equation}\nonumber
    A_i(z^{-1})y_i(t)=B_i(z^{-1})u(t-1),\quad (i=1,2)
\end{equation}
where we assume all with 3 unknown parameters,
\begin{equation}\nonumber
\begin{split}
   A_1(z^{-1})=1+\sum_{k=1}^{3}{a}_{1,k}z^{-k},\quad
   A_2(z^{-1})=1+\sum_{k=1}^{3}{a}_{2,k}z^{-k}.\\
   B_1(z^{-1})=z^{-1}\sum_{k=0}^{2}{b}_{1,k}z^{-k},\quad
  B_2(z^{-1})=z^{-1}\sum_{k=0}^{2}{b}_{2,k}z^{-k}.
\end{split}
\end{equation}
Applying a least square method we obtain
\begin{equation}\nonumber
\begin{split}
  &y(t) -\tilde{y}(t)=\hat{F} u(t) \\
    &= \bmat \frac{0.2901 + 0.7977  z^{-1} + 0.4339 z^{-2} }{1 + 0.2137 z^{-1}-0.02525z^{-2} -0.05393 z^{-3}  } \\ \frac{ 0.1302  + 0.9191 z^{-1} - 0.6492 z^{-2}}{1 - 0.1363   z^{-1}  - 0.1090 z^{-2}  - 0.05338 z^{-3} }\emat u(t-1).
\end{split}
\end{equation}
The corresponding Bode diagrams are shown in Fig.~\ref{FigExmp2F1} and in Fig.~\ref{FigExmp2F2}.

\begin{figure}
      \centering
      \includegraphics[scale=0.6]{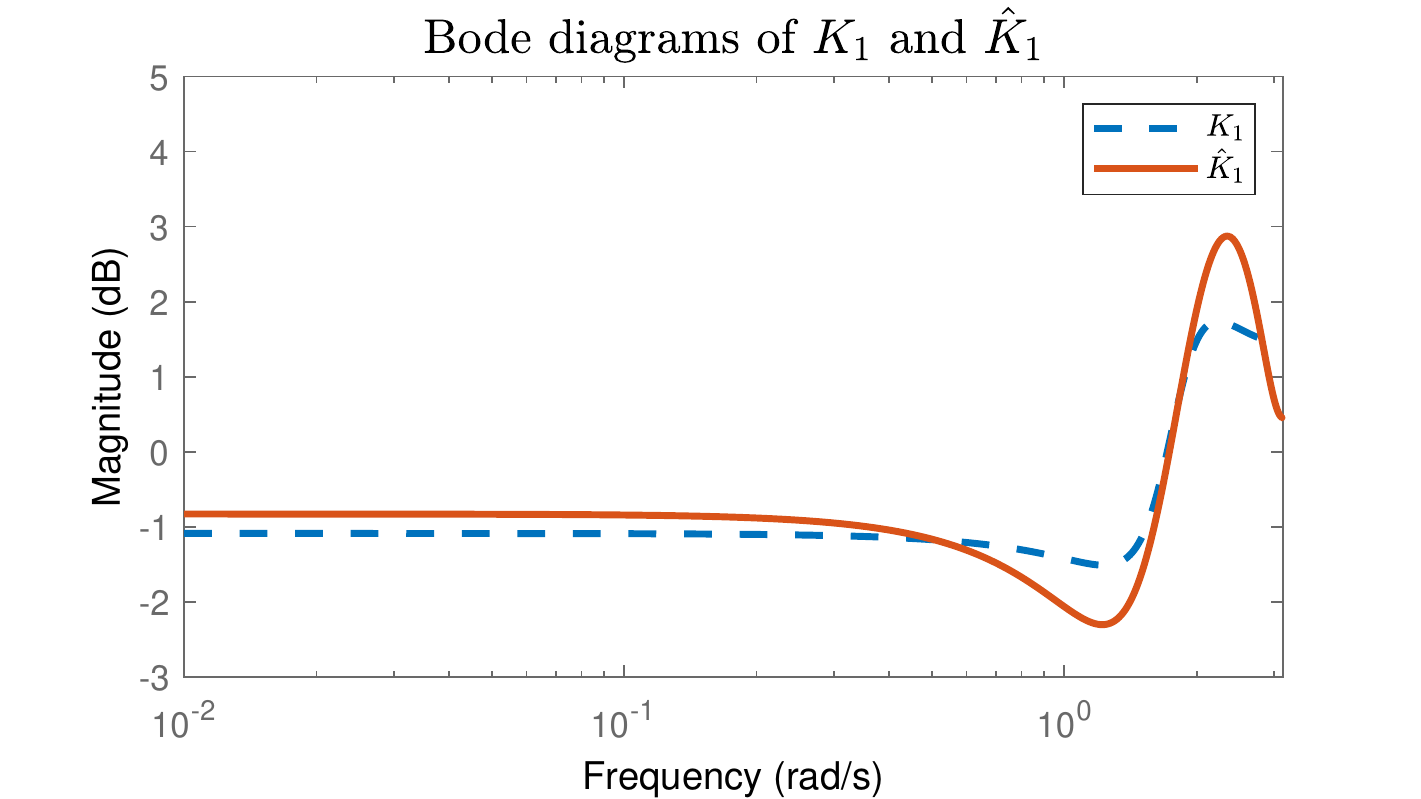}
      \caption  {Bode diagrams of $K_1$ and $\hat{K}_1$ in example 2.}
      \label{FigExmp2K1}
\end{figure}

\begin{figure}
      \centering
      \includegraphics[scale=0.6]{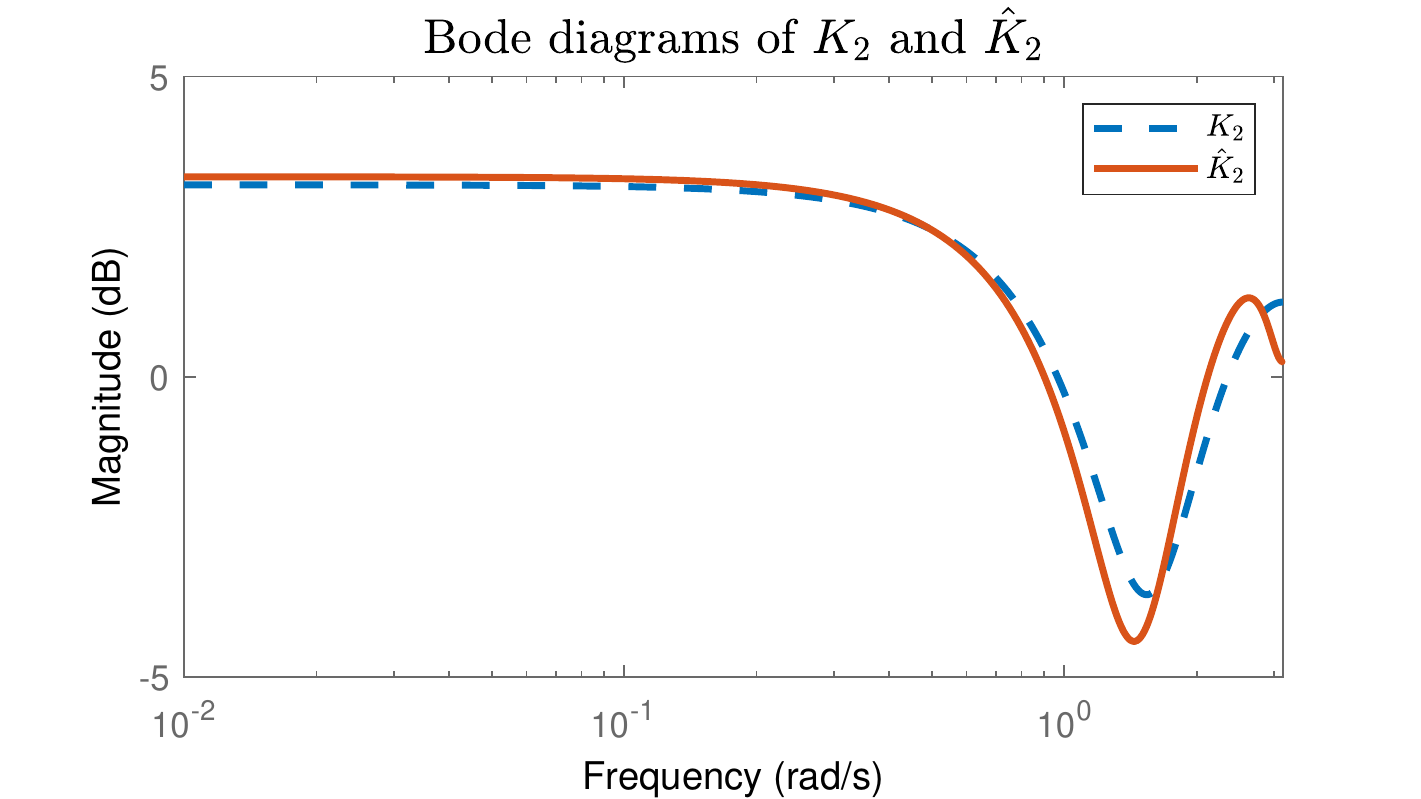}
      \caption {Bode diagrams of $K_2$ and $\hat{K}_2$ in example 2.}
      \label{FigExmp2K2}
\end{figure}

Then we estimate $K(z)$ from \eqref{tildey} by the same procedure  we used to estimate $W(z)$ in \eqref{W1W2}. Suppose
$$
 A_i(z^{-1})\tilde{y}(t)=B_i(z^{-1})e(t), \quad (i=1,2)
$$
with (since $K$ is normalized)
\begin{equation}\nonumber
\begin{split}
  A_1(z^{-1})=1+\sum_{k=1}^{3}{a}_{1,k}z^{-k},\quad
  A_2(z^{-1})=1+\sum_{k=1}^{3}{a}_{2,k}z^{-k}.\\
  B_1(z^{-1})=1+\sum_{k=1}^{3}{b}_{1,k}z^{-k},\quad
  B_2(z^{-1})=1+\sum_{k=1}^{3}{b}_{2,k}z^{-k}.
\end{split}
\end{equation}
and obtain
$$
\hat{K}(z)=\bmat  \frac{1+ 0.4940  z^{-1} + 0.2391z^{-2} + 0.1936z^{-3}}{1 + 0.7235 z^{-1} + 0.3215z^{-2} +0.07442 z^{-3}}\\
    \frac{1+ 0.5175  z^{-1} + 0.3272 z^{-2}  + 0.03482z^{-3} }{1  + 0.4528 z^{-1} -0.0283z^{-2} +0.07029 z^{-3}} \emat,
$$
whose corresponding Bode diagrams are in Fig.~\ref{FigExmp2K1} and Fig.~\ref{FigExmp2K2}.
 Here we obtain reasonable estimates of both $K_1$   and $K_2$.

\subsection{Example 3}
\begin{figure}
      \centering
      \includegraphics[scale=0.6]{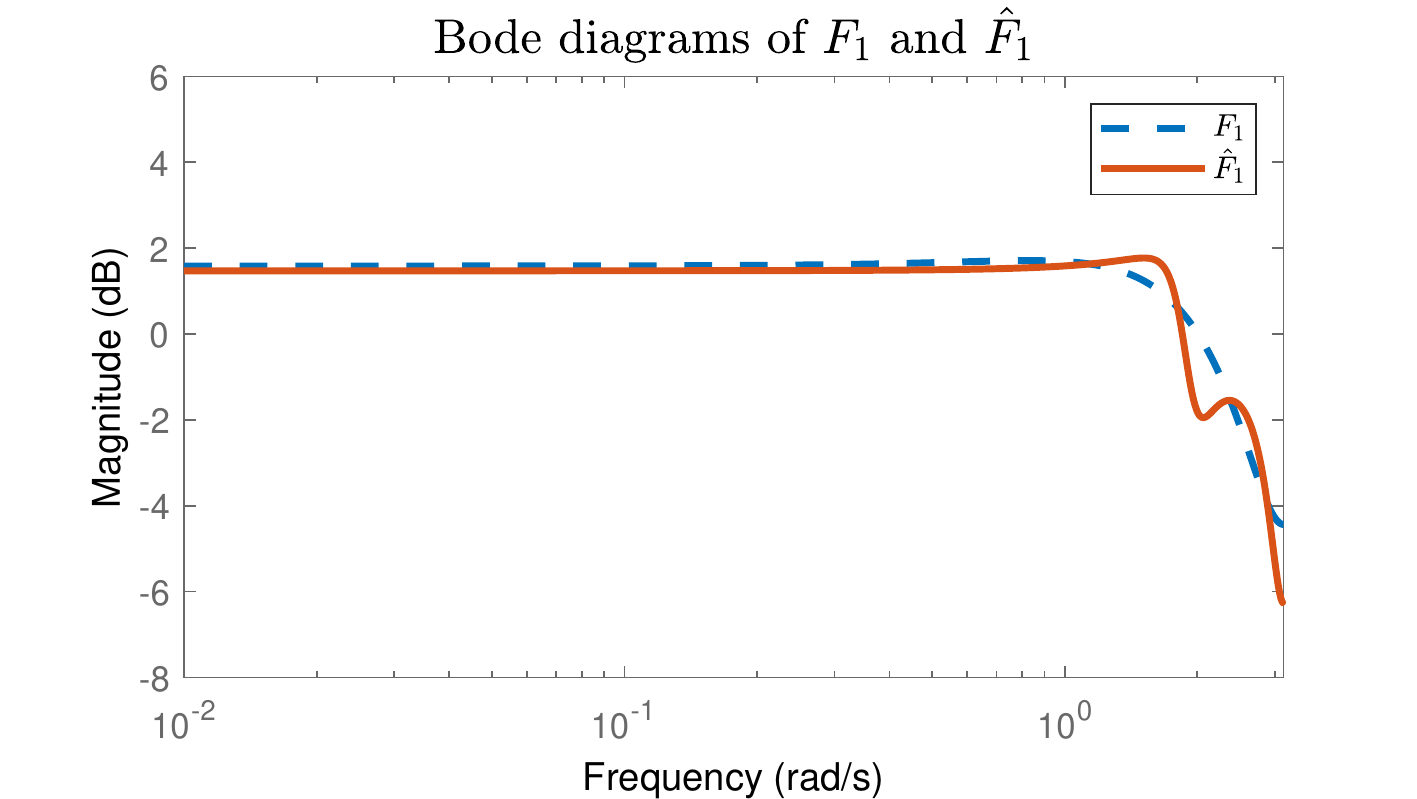}
      \caption{  Bode diagrams of $F_1$ and $\hat{F}_1$ in example 3}
      \label{FigExmp2F1new}
\end{figure}

\begin{figure}
      \centering
      \includegraphics[scale=0.6]{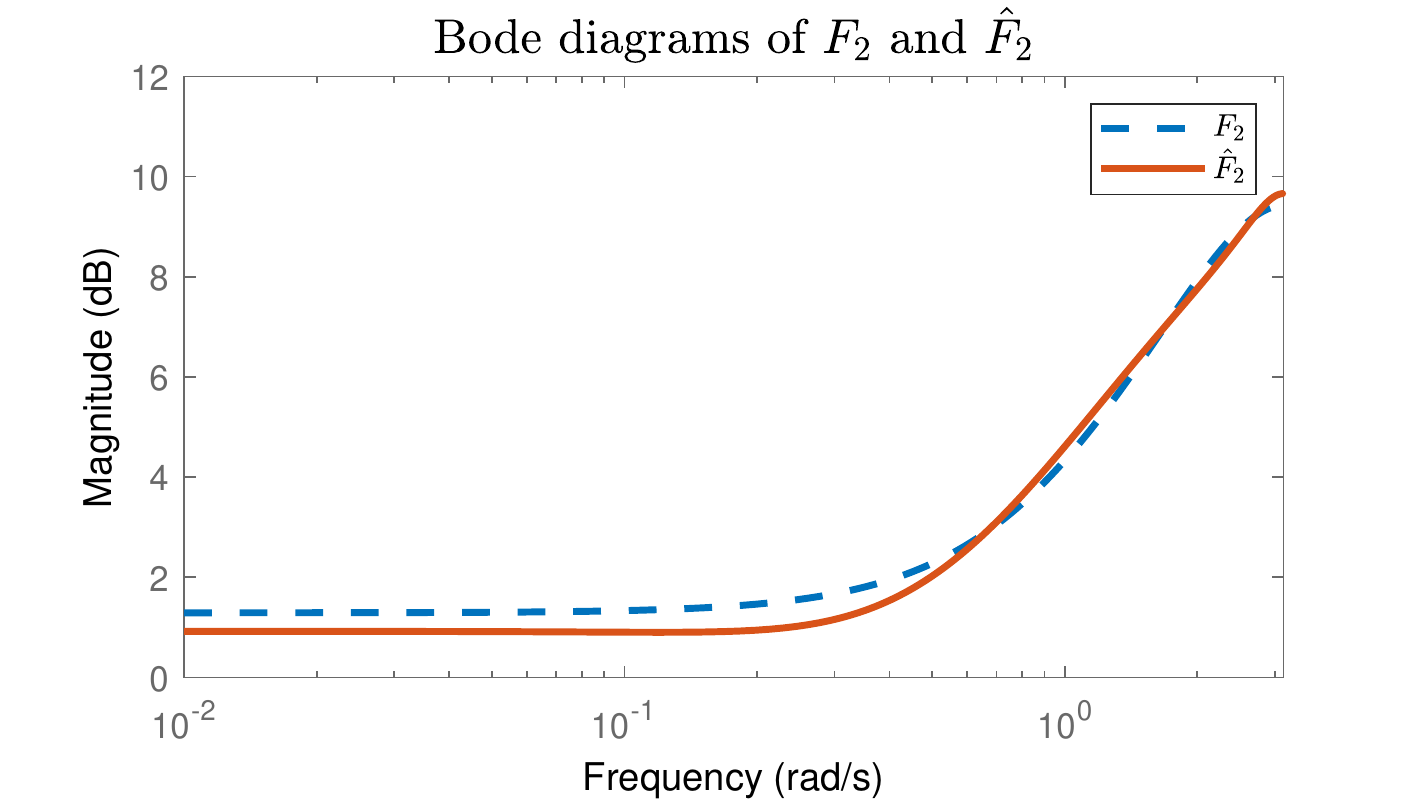}
      \caption{  Bode diagrams of $F_2$ and $\hat{F}_2$ in example 3}
      \label{FigExmp2F2new}
\end{figure}

Again, we generate  a scalar white noise input $u$ independent of $e$ and identify a two-dimensional system   \eqref{WithInput}, as in the previous subsection. The  true system,  is described by
\begin{equation}\nonumber
\begin{split}
   F(z) &= z^{-1} \bmat 1 + 0.3z^{-1}-0.1z^{-2}\\
                 2 - 0.9z^{-1}+0.06z^{-2}\emat, \\
   K(z) &= \bmat \frac{1-0.9z^{-1}+0.2z^{-2}}{1+0.3z^{-1}+0.4z^{-2}}\\ \frac{1-0.1z^{-1}+0.4z^{-2} }{1-0.6z^{-1}+0.1z^{-2}} \emat.
\end{split}
\end{equation}
The simulation  is run with $u$ and $e$ two independent scalar white noises of variances $2$ and $1$.
We generate a two-dimensional  time series of $N=500$ data points $\{\bar y_i(t); t=1,\ldots,N,\, i=1,2 \}$, and suppose we also measure  the input time series of $u$. Firstly, we estimate $F(z)$ by fitting the deterministic relation  $ y(t) =F(z) u(t)$ rewritten as
\begin{equation}\nonumber
    A_i(z^{-1})y_i(t)=B_i(z^{-1})u(t-1),\quad (i=1,2)
\end{equation}
where the polynomials are chosen of degree 3, i.e.
\begin{equation}\nonumber
\begin{split}
  A_1(z^{-1})&=1+\sum_{k=1}^{3}{a}_{1,k}z^{-k},\quad
  A_2(z^{-1})=1+\sum_{k=1}^{3}{a}_{2,k}z^{-k}.\\ 
  B_1(z^{-1})&=\sum_{k=0}^{3}{b}_{1,k}z^{-k},\quad
  B_2(z^{-1})=\sum_{k=0}^{3}{b}_{2,k}z^{-k}.
\end{split}
\end{equation}
Applying a least square method we obtain
\begin{equation}\nonumber
\begin{split}
  &y(t) -\tilde{y}(t)=\hat{F} u(t) \\
    &= \bmat \frac{0.9807 + 1.353  z^{-1} + 1.114 z^{-2} + 0.5196 z^{-3}}{1 + 1.064 z^{-1}+0.903 z^{-2} + 0.3815 z^{-3}  } \\ \frac{ 1.991 - 1.831  z^{-1} - 0.09642  z^{-2} + 0.2789z^{-3}}{1 - 0.4595 z^{-1}  - 0.2792 z^{-2} + 0.04673 z^{-3} }\emat u(t-1).
\end{split}
\end{equation}
The corresponding Bode graphs are shown in Fig.~\ref{FigExmp2F1new} and Fig.~\ref{FigExmp2F2new}. In Fig.~\ref{FigExmp2F1new}, the Bode graph of $\hat{F}_1$ shows some overfitting  since the order of $F$  is somewhat far from the true order (in fact $A_1=1$ with order $0$, but we suppose a degree of $3$. Assuming we know  the orders of $A_1$, $B_1$, we get the estimate
$$
    \hat{F}_1'=  0.9802z^{-1} + 0.314 z^{-2} - 0.09327z^{-3},
$$
which is closer to $F_1$.

\begin{figure}
      \centering
      \includegraphics[scale=0.6]{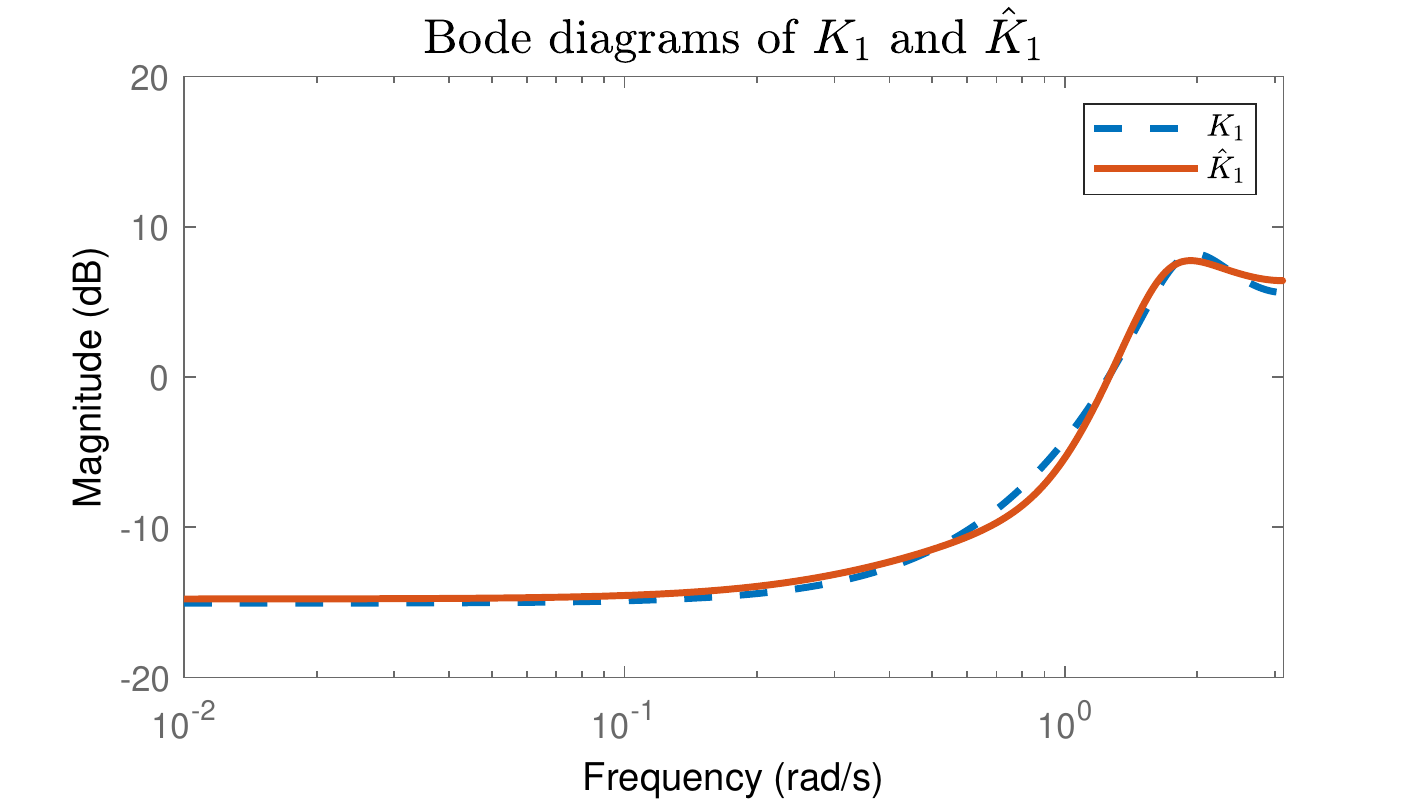}
      \caption{ Bode diagrams of $K_1$ and $\hat{K}_1$ in example 3.}
      \label{FigExmp2K1new}
\end{figure}

\begin{figure}
      \centering
      \includegraphics[scale=0.6]{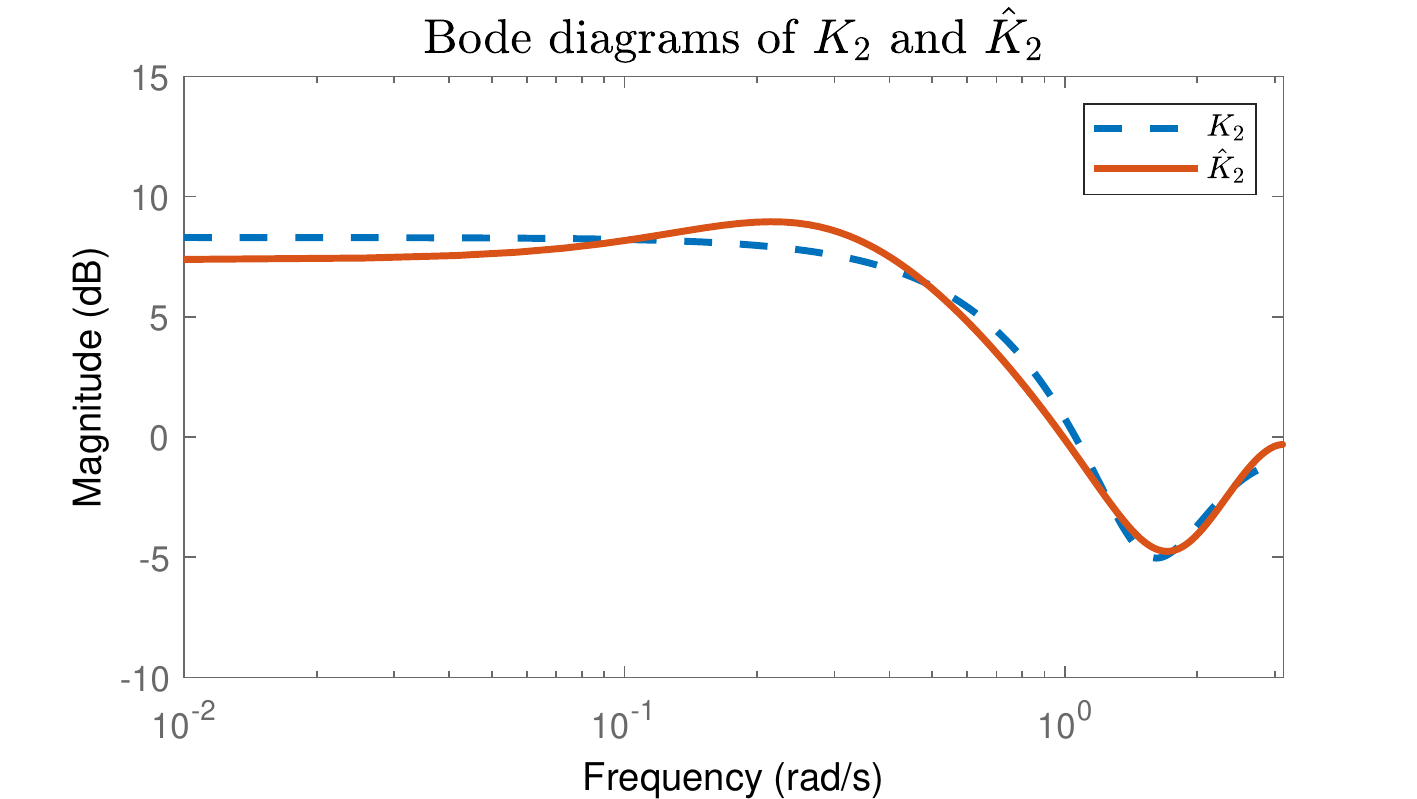}
      \caption {Bode diagrams of $K_2$ and $\hat{K}_2$ in example 3.}
      \label{FigExmp2K2new}
\end{figure}

Then we estimate $K(z)$ from \eqref{tildey} by the same procedure  we used to estimate $W(z)$ in \eqref{W1W2}. Suppose
$$
 A_i(z^{-1})\tilde{y}(t)=B_i(z^{-1})e(t), \quad (i=1,2)
$$
with (since $K$ is normalized)
\begin{equation}\nonumber
\begin{split}
  A_1(z^{-1})=1+\sum_{k=1}^{3}{a}_{1,k}z^{-k},\quad
  A_2(z^{-1})=1+\sum_{k=1}^{3}{a}_{2,k}z^{-k}.\\
  B_1(z^{-1})=1+\sum_{k=1}^{3}{b}_{1,k}z^{-k},\quad
  B_2(z^{-1})=1+\sum_{k=1}^{3}{b}_{2,k}z^{-k},
\end{split}
\end{equation}
 obtaining
$$
\hat{K}(z)=\bmat \frac{1- 1.481 z^{-1} + 0.9142  z^{-2}  - 0.2516z^{-3}}{1 - 0.2452 z^{-1} + 0.3701 z^{-2} - 0.1293z^{-3}}\\
    \frac{1 - 0.8098z^{-1} + 0.2342 z^{-2} - 0.2265z^{-3} }{1 - 1.28 z^{-1}  + 0.2189 z^{-2}  + 0.1462z^{-3}} \emat,
$$
 the corresponding   Bode diagrams are in Fig.~\ref{FigExmp2K1new} and Fig.~\ref{FigExmp2K2new}.

All the simulation examples show that the transfer functions of the rank-deficient structure can be identified from standard identification algorithms with rather good  results. Of course, with a prior knowledge of the orders of the transfer functions,   the identification results will be closer to the true functions.

\section{Conclusions}\label{secCon}
We have shown that a rank-deficient process admits a special feedback structure with a deterministic feedback channel which can be used to split the identification in two steps, one of which can be based on standard PEM algorithms while the other is based on a deterministic least squares fit. Simulations show that standard identification algorithms can be easily applied to identify the transfer functions of this structure.

\bibliography{ifacconf}             

\end{document}